\documentstyle[fleqn,twoside,gc]{article}

\def\be{\begin{equation}}
\def\ee{\end{equation}}
\def\bea{\begin{eqnarray}}
\def\eea{\end{eqnarray}}
\def\nn{\nonumber \\}
\def\e{{\rm e}}
\def\SEH{S_{\rm EH}}
\def\SGH{S_{\rm GH}}
\def\AdS5{{{\rm AdS}_5}}
\def\S4{{{\rm S}_4}}

\def\gfr{{g_{(4)}}}

\def\wlBox{\mbox{$\widetilde{\DAL}$}}

\heads{K.E. Osetrin}{Dilatonic Quantum-Induced Two-Brane Worlds}

\bls{1.02}
\begin{document}
\twocolumn[ \Arthead{8}{2002}{1/2 (29/30)}{87}{90}

\bigskip
\Title{DILATONIC QUANTUM-INDUCED TWO-BRANE WORLDS}

\Author{K.E. Osetrin}{Tomsk State Pedagogical University, Tomsk, Russia}

\Rec{28 July 2001}

\Abstract
{The 5d dilatonic gravity action with surface counterterms, motivated by
the AdS/CFT correspondence and with contributions of brane quantum CFTs is
considered with an AdS-like bulk. The role of quantum brane CFT consists in
inducing complicated brane dilatonic gravity. For exponential bulk
potentials, a number of AdS-like bulk spaces is found in an analytical form.
The corresponding flat or curved (de Sitter or hyperbolic) dilatonic
two-branes are created.}


] 

\email 1 {osetrin@tspu.edu.ru\\
This contribution is based on a common work with S. Nojiri and
S.D. Odintsov [\,1\,]}

\section{Introduction}

The recent booming activity in brane-world studies is caused
by several reasons. First, gravity on a 4d brane embedded in
a higher-dimensional AdS-like Universe may be localized \cite{RS1,RS2}.
Second, there appears a way to resolve the mass hierarchy problem \cite{RS1}.

An essential element of brane-world models is the presence of two free
parameters in the theory (the bulk cosmological constant and the brane
tension, or brane cosmological constant). The role of the brane cosmological
constant is to fix the position of the brane in terms of tension (that is,
why the brane cosmological constant and the brane tension are almost the same
thing).  Being completely consistent and mathematically reasonable, such a
way of doing things may look incompletely satisfactory.  Indeed, the
physical origin (and prediction) of brane tension in terms of some dynamical
mechanism may be required.

The ideology may be different, in the spirit of Refs.\,\cite{NOZ, HHR,cosm}.
One considers the addition of surface counterterms to the original action on
an AdS-like space. These terms are responsible for making the variational
procedure well-defined (in Gibbons-Hawking's spirit) and for eliminating the
leading divergences of the action. The brane tension is no more considered
as a free parameter but is fixed by the condition of finiteness of
spacetime when the brane goes to infinity. Of course, leaving the theory in
such a form would rule out a possible existence of consistent brane-world
solutions. Fortunately, other parameters contribute to brane tension. If one
assumes that there is a quantum CFT living on the brane (which is closer
to the spirit of AdS/CFT correspondence \cite{AdS}), then such a CFT
produces a conformal anomaly (or an anomaly-induced effective action).  This
contributes to the brane tension. As a result, a dynamical mechanism of
getting a brane world with a flat or curved (de Sitter or Anti-de Sitter)
brane appears. In other words, the brane world is a consequence of the
presence of matter on the brane! For example, the sign of conformal anomaly
terms for usual matter is such that, in the one-brane case, the de Sitter
(ever-expanding, inflationary) Universe is a preferable solution to the
brane equation\footnote{A similar mechanism for anomaly driven inflation in
the usual 4d world has been invented by Starobinsky \cite{SMM} and generalized
for the presence of a dilaton in Refs.\,\cite{Brevik}}.

The scenario of Refs.\,\cite{NOZ,HHR} may be extended to the presence
of dilaton(s), as was done in \Ref{NOO}, or to the formulation of
quantum cosmology in a Wheeler-De Witt form \cite{ANO}. Then the whole
scenario looks even more related to the AdS/CFT correspondence, since
dilatonic gravity appears naturally as the bosonic sector of 5d gauged
supergravity.  Moreover, there appears an extra prize in the form of
dynamical determination of the 4d boundary value of the dilaton. In
\cite{NOO}, a quantum dilatonic one-brane Universe has been presented,
with a possibility of getting an inflationary, hyperbolic or flat brane with
dynamical determination of the brane dilaton. An interesting question is
related to generalization of such a scenario in dilatonic gravity for the
multi-brane case. This will be discussed here.

In the next section we present the general action of 5d dilatonic
gravity with surface counterterms and a quantum brane CFT
contribution. This action is convenient for describing
brane-worlds where the bulk is an AdS-like spacetime. There could
be one or two (flat or curved) branes in the theory. As was
already mentioned, the brane tension is not fixed in our
approach, instead, the effective brane tension is induced by
quantum effects. An explicit analytical solution of the bulk
equation for a number of exponential bulk potentials is
presented. It is interesting that quantum-created branes can be
flat, de Sitter (inflationary), or hyperbolic. The role of
quantum brane matter corrections in getting such branes is
extremely important.  Nevertheless, there are a few particular
cases where such branes appear on the classical level, i.e.,
without quantum corrections.

In most cases, as usually occurs in AdS dilatonic gravity,
the solutions contain a naked singularity. However, in other cases
the scalar curvature is finite, and there is a horizon.  The corresponding
4d branes may be interpreted as wormholes.

\section{Dilatonic gravity action with brane quantum corrections}

Let us present the initial action for dilatonic AdS gravity under
consideration. The (Euclidean) AdS metric has the following form:
\be
\label{AdS}
     ds^2=dz^2 + \e^{2\tilde A(z)}\sum_{i,j=1}^4\hat g_{ij}dx^i dx^j.
\ee
Here $\hat g_{ij}$ is the metric of the Einstein manifold defined by
$r_{ij}=k\hat g_{ij}$, where $r_{ij}$ is the Ricci tensor constructed from
$\hat g_{ij}$ and $k$ is a constant.  One can consider two copies of the
regions given by $z<z_0$ and glue two regions  putting a brane at $z=z_0$.
More generally, one can consider two copies of regions
$\tilde z_0<z<z_0$ and glue the regions putting two branes
at $z=\tilde z_0$ and $z=z_0$. Hereafter we call the brane at $z=\tilde z_0$
as an ``inner' brane and that at $z=z_0$ as ``outer'' brane.

One chooses the 4-dimensional boundary metric as
\be                                               \label{tildeg}
    \gfr_{\mu\nu}=\e^{2A}\tilde g_{\mu\nu},
\ee
and a tilde specifies quantities connected with $\tilde g_{\mu\nu}$.

Let us consider the case where there are two branes at
$z=\tilde z_0$ and $z=z_0$:
\bearr                                  \label{Stotal2}
    S_{\rm two\ branes}=S+ \tilde\SGH + 2 \tilde S_1 + \tilde W,
\\ \lal \label{SEHib}
    \tilde\SGH={1 \over 8\pi G}\int d^4 x
        \sqrt{\gfr}\ \nabla_\mu n^\mu,
\\ \lal \label{S1b}
    \tilde S_1= {1 \over 16\pi G l}\int d^4 x \sqrt{\gfr}\left(
        {6 \over l} + {l \over 4}\Phi(\phi)\right),
\\ \lal \label{Wb}
    \tilde W= \tilde b \int d^4x \sqrt{\widetilde g}\widetilde F A
\nnn
    + \tilde b' \int d^4x\sqrt{\widetilde g}
    \biggl\{A \biggl [ 2 {\wlBox}^2
        +\widetilde R_{\mu\nu}\widetilde\nabla_\mu\widetilde\nabla_\nu
\nnn
     - {4 \over 3}\widetilde R \wlBox^2
    + {2 \over 3}(\widetilde\nabla^\mu \widetilde R)\widetilde\nabla_\mu
         \biggr]A
          + \left(\widetilde G - {2 \over 3}\wlBox \widetilde R
                \right)A \biggr\}
\nnn
    -{1 \over 12}\left[ \tilde b''+ {2 \over 3}(\tilde b
                + \tilde b')\right]
    \int d^4x \sqrt{\widetilde g} \biggl [ \widetilde R
\nnn
    - 6\wlBox A - 6 (\widetilde\nabla_\mu A)(\widetilde \nabla^\mu A)
            \biggr]^2
\nnn
    + \tilde C \int d^4x \sqrt{\widetilde g}
    A \phi \biggl [ {\wlBox}^2
        + 2\widetilde R_{\mu\nu}\widetilde\nabla_\mu\widetilde\nabla_\nu
\nnn  \cm
        - {2 \over 3}\widetilde R \wlBox^2
      + {1 \over 3}(\widetilde\nabla^\mu \widetilde R)\widetilde\nabla_\mu
        \biggr] \phi
\ear
with $N$ scalar, $N_{1/2}$ spinor, $N_1$ vector fields, $N_2$ ($=0$ or
$1$) gravitons and $N_{\rm HD}$ higher-derivative conformal scalars;
$b$, $b'$ and $b''$ are
\bear                                                          \label{bs}
      b\eql {N +6N_{1/2}+12N_1 + 611 N_2 - 8N_{\rm HD}\over 120(4\pi)^2},
\nn
      b' \eql -{N+11N_{1/2}+62N_1 + 1411 N_2 -28 N_{\rm HD}
            \over 360(4\pi)^2}\ ,
\nn
    b''\eql 0\ .
\eea
For an introduction to the anomaly-induced EA see \cite{BOS}. We should note
that the relative sign of $\tilde S_1$ is different from $S_1$. The
parameters $\tilde b$, $\tilde b'$, $\tilde b''$ and $\tilde C$ correspond
to the matter which may be different from the outer brane one on the inner
brane as in (\ref{bs}).  Hence, the situation with different CFTs on the
branes may be considered.  Having the action, one can study its
dynamics.

Let us start the consideration of field equations for a two-brane model.
First of all,  one defines a new coordinate $z$ by
\beq                                                       \label{c2b}
    z=\int dy\sqrt{f(y)},
\eeq
and solves $y$ with respect to $z$. Then the warp factor is $\e^{2\hat
A(z,k)}=y(z)$. Here one assumes the 5-dimensional space-time metric as
follows:
\beq                                                        \label{DP1}
    ds^2=f(y)dy^2 + y\sum_{i,j=1}^4\hat g_{ij}(x^k)dx^i dx^j.
\eeq
Using the substitution $dz=\sqrt{f}dy$ and choosing $l^2\e^{2\hat
A(z,k)}=y(z)$, one has metric in the form
\bearr                                                       \label{metric1}
   ds^2=dz^2 + \e^{2A(z,\sigma)}\tilde g_{\mu\nu}dx^\mu dx^\nu\ ,
\nnn \cm
    \tilde g_{\mu\nu}dx^\mu dx^\nu\equiv l^2\left(d \sigma^2
        + d\Omega^2_3\right)\ .
\ear
Here $d\Omega^2_3$ is the metric of a 3-dimensional
unit sphere. Then for the unit sphere ($k=3$)
\be                                                  \label{smetric}
    A(z,\sigma)=\hat A(z,k=3) - \ln\cosh\sigma\ ,
\ee
for flat Euclidean space ($k=0$)
\be                                                  \label{emetric}
    A(z,\sigma)=\hat A(z,k=0) + \sigma\ ,
\ee
and for the unit hyperboloid ($k=-3$)
\be                                                  \label{hmetric}
    A(z,\sigma)=\hat A(z,k=-3) - \ln\sinh\sigma\ .
\ee
We now identify $A$ and $\tilde g$ in (\ref{metric1}) with those in
(\ref{tildeg}). Then we find $\tilde F=\tilde G=0$,
$\tilde R=6/l^2$, etc.

Our choice for the dilaton and bulk potential admitting an analytical
solution is
\bear                                                       \label{assmp1}
    \phi(y) \eql p_1\ln \left(p_2 y\right),
\\   \label{assmp2}
    \Phi(\phi) \eql c_0 + c_1 \e^{a\phi} + c_2\e^{ 2a\phi},
\ear where $a$, $p_1$, $p_2$, $c_0$, $c_1$, $c_2$ are some
constants. When $p_1=\pm 1 /\sqrt{6}$, we find that $f(y)$
identically vanishes. Therefore we should assume $p_1\neq \pm
1/\sqrt{6}$. Then we find the following set of exact bulk
solutions (for all cases $c_0=-{12/l^2}$): \bear
\nq
\label{case1} \mbox{case 1:}
    \quad && c_1={6kp_2p_1^2 \over 3 - 2p_1^2}\ ,\quad
        c_2=0,  \quad a=-{1 \over p_1}\ ,
\nnn
    f(y)={3- 2p_1^2 \over 4ky}\ ,\quad p_1\neq \pm \sqrt{6};
\\ \label{case2} \nq
\mbox{case 2:}
    \quad && c_1=-6kp_2,\quad
        a=\pm{1 \over \sqrt{3}}\ ,\quad p_1=\mp\sqrt{3},
\nnn
    f(y)={3 \over {2c_2 \over p_2^2} - 4ky};
\\ \label{case3} \nq
\mbox{case 3}
    \quad && c_2=3kp_2,\quad
    a=\pm{1 \over \sqrt{3}}\ ,\quad p_1=\mp{\sqrt{3} \over 2},
\nnn
    f(y)={21\sqrt{p_2} \over 8\sqrt{y}\left(c_1y
                    + 7k\sqrt{p_2y}\right)} \ .
\ear
The equation for the case $k=0$ is identical to that of the outer brane.

As an example, we consider case 1 in (\ref{case1}). Since $f(y)$ should
be positive (we should also note $y>0$), one gets
\beq                                                    \label{qq}
    q^2\equiv {4k \over 3 - 2p_1^2}>0 \qquad (q>0)\ .
\eeq
In (\ref{qq}), we can also consider the limit of $k\to 0$ keeping
$q$ finite, i.e., $p_1^2\to {3 \over 2}$.

When $k\neq 0$, from the equations of motion we get:
\bear                                                   \label{eq2c11}
    -8b'\eql F_1(y_0)\equiv {3y_0 \over 32\pi G}\left(
        q y_0^{1/2} - \frac{y_0}{l}
                - {q^2p_1^2l\over 8}\right)
\nn
    \eql  - {3 \over 16\pi G}{y_0 \over 2l}
    \biggl( y_0^{1/2} - {1 + \sqrt{1 - {p_1^2 l^2/2}}\over 2}q\biggr)
\nnn \
      \times \biggl( y_0^{1/2} - {1 - \sqrt{1 - {p_1^2 l^2/2}}
                \over 2}q\biggr)
\\  \label{eq2pc1}
    0\eql  - {q \over 8\pi G}y_0^{5/2}
            + {9lp_1 q^2 \over 8\pi G}y_0 + 6C\phi_0\ .
\eea
The brane equations are
\bear                               \label{eq2c11in}
    8\tilde b'\eql F_1(y_0),
\\                                      \label{eq2pc1in}
    0 \eql  {p_1 q \over 8\pi G}y_0^{3 \over 2}
        - {9lp_1 q^2 \over 8\pi G}y_0 + 6\tilde C\tilde\phi_0\ .
\ear
Since $p_2$ is absorbed into the definition of $q$ in (\ref{eq2c11}) and
(\ref{eq2c11in}), \eqs (\ref{eq2pc1}) and (\ref{eq2pc1in}) can be regarded
as the equation which determines $p_2$, or
$(\phi_0/p_1)\ln (p_2 y_0)$ and $(\tilde \phi_0/p_1)=\ln(p_2 y_0)$.
We now investigate the properties of $F_1(y_0)$ as a function of $y_0$. The
asymptotic behaviours are given by
\bear                                                         \label{F1}
    F_1(y_0)&\stackrel{y_0 \to +0}{\longrightarrow}&
        - {3 \over 16\pi G}\cdot{p_1^2  q^2 l \over 16}y_0 < 0,
\\ \label{F2}
    F_1(y_0)&\stackrel{y_0 \to +\infty }{\longrightarrow}&
            - {3 \over 16\pi G}\cdot {1 \over 2l}y_0^2 < 0\ .
\ear
Since
\beq                                        \label{F3}
    F'_1(y_0)= {3 \over 16\pi G}\left( {3q \over 4}y^{1/2}
            - {1 \over l}y_0 - {p_1^2 q^2 l \over 16}\right)\ ,
\eeq
$F_1(y_0)$ has extrema when
\beq                                                \label{F4}
    0=y_0 - {3ql \over 4}y_0^{1/2}+ {p_1^2 q^2 l^2 \over 16}\ ,
\eeq
whose solutions are given by
\beq                                                  \label{F5}
    y_0^{1/2}=y_\pm^{1/2} \equiv {3ql \over 8}\biggl(1\pm \sqrt{1
                - {4p_1^2 \over 9}}\biggr)\ .
\eeq
Therefore if
\beq                                                      \label{F6}
        |p_1|>{3/2}\ ,
\eeq
\eq (\ref{F4}) has no solution and $F_1(y_0)$ is a monotonically decreasing
function of $y_0$. Then \eqs (\ref{F1}) and (\ref{F2}) tell that there is no
solution of the brane equation (\ref{eq2c11}) for negative $b'$ in the case
(\ref{F6}).  On the other hand, when
\beq                                                  \label{F6b}
    |p_1|<{3 /2}\ ,
\eeq
 substituting (\ref{F5}) into the expression for $F_1(y_0)$ in
(\ref{eq2c11}), one gets
\bear
    F_1(y_\pm)\eql {3 \over 16\pi G}{3^4 q^4 l^3 \over 2\cdot 8^4}
        \biggl(\sqrt{1-{4p_1^2 \over 9}} \pm 1\biggr)
\nnn \cm \label{F7}
        \times\biggl(\sqrt{1-{4p_1^2 \over 9}} \mp {1 \over 3}\biggr)\ .
\ear
Then we find that $y_0=y_+$ corresponds to a maximum of $F_1(y_0)$. The
maximum is positive, $F_1(y_+)>0$, if $\sqrt{1- 4p_1^2/9}-1/3 >0$, that is,
\beq                        \label{F8}
        p_1^2<2 .
\eeq
In the case (\ref{F8}), if
\beq                            \label{F9}
    F_1(y_+)\geq -8b',
\ee
\eq (\ref{eq2c11}) has a solution, i.e., there can be a brane. We can also
consider an inner brane, which lies at $y=y_1<y_0$. For the inner brane, the
relative sign of $\tilde b'$ and $b'$ changes in the equation
corresponding to (\ref{eq2c11}).  Then, if
\beq                                        \label{F10}
        F_1(y_-)\leq 8\tilde b',
\eeq
there can be an inner brane. Then, if both (\ref{F9}) and (\ref{F10}) hold,
we can have a two-brane dilatonic solution. For such a solution, there might
be, in general, a problem in the consistency between (\ref{eq2pc1}) and
(\ref{eq2pc1in}). Taken together, (\ref{eq2pc1}) and (\ref{eq2pc1in}),
can be regarded as the equations which determine $p_1$ and $p_2$ (we
should note that $p_2$ is implicitly contained in $\phi_0$ and
$\tilde\phi_0$).  In the classical limit, where $C=0$, the terms containing
$p_2$ (or $\phi_0$ and $\tilde\phi_0$) disappear. Then it seems to be
non-trivial whether or not there is any solution which satisfies both
(\ref{F9}) and (\ref{F10}).

We now consider the classical limit in case $k\neq 0$ , where
$b'=C=\tilde b'=\tilde C=0$, and (\ref{eq2c11}) and (\ref{eq2c11in}) become
identical. Then the solutions to \eqs (\ref{eq2c11}) and
(\ref{eq2c11in})  are given by
\beq                                                     \label{cc1a}
    y_0^{1 \over 2} = \biggl(1 \pm \sqrt{1 - {p_1^2 \over 2}}
                \biggr){ql \over 2}\ .
\eeq
Since both solutions are positive, we can regard the smaller one (the $-$
sign in (\ref{cc1a})) as the one corresponding to the inner brane and
and larger one ($+$ sign) to the outer brane.  On the other hand,
\eqs (\ref{eq2pc1}) and (\ref{eq2pc1in}) have the following form:
\beq                                                          \label{cc1b}
    0={p_1 q^2 l y_0 \over 2}\biggl(17\mp \sqrt{1
            - {p_1^2 \over 2}}\biggr) .
\eeq
In (\ref{cc1b}), the upper sign ($-$) corresponds to the outer brane and the
lower one ($+$) to the inner brane. We should note that there is no
solution, except the trivial one, for the inner brane. This seems to tell us
that we need a quantum correction from brane matter in order to obtain the
two-brane dilatonic inflationary universe where the observable world can be
associated with one of inflationary branes.

When $k=0$ in case 1, as discussed before, if $q$ is finite,
\beq                                                         \label{c1k0i}
    p_1^2\to 3/2.
\eeq
Then we get the following equation:
\beq                                               \label{c1k0ii}
    0=y_0 - ql y_0^{1/2} + {3q^2l^2 \over 16},
\eeq
which has two solutions:
\beq                                              \label{c1k0iii}
    y_0^{1/2}= {3ql \over 4}\ ,\quad {ql \over 4}.
\eeq
These two solutions might be regarded as two-brane solutions. On the other
hand, the form of the equation of motion in case $k=0$ is identical
to that of $k\neq 0$ in (\ref{eq2pc1}), which can be solved with respect
to $\phi_0$ or $p_2$ in the one-brane solution.  However, in case $k=0$, the
value of $p_1$ is fixed by (\ref{c1k0i}). In the classical limit of the
case $k=0$, the equation of motion \cite{NOO1} can be rewritten in the form
of (\ref{c1k0ii}), and there appear the solutions (\ref{c1k0iii}).
The brane equation of motion \cite{NOO1} has the form (\ref{cc1b}), but
\eq (\ref{c1k0i}) does not satisfy (\ref{cc1b}).  This demonstrates the role
of quantum effects in the realization of a dilatonic inflationary
two-brane-world Universe.

Solutions 2 and 3 may be considered in a similar way, for details
see \cite{NOO1}.

\section{Discussion}

In summary, we have presented a generalization of a quantum dilatonic
brane world\cite{NOO} where the brane is flat, spherical (de Sitter) or
hyperbolic, and it is induced by the quantum effects of CFT living on the
brane.  In this generalization one may have two-brane-worlds or even
multi-brane-worlds what proves a general nature of the scenario suggested in
Refs.\,\cite{HHR,NOZ} where, instead of an arbitrary brane tension added by
hand, the effective brane tension is produced by boundary quantum fields.
What is more interesting, the bulk solutions have an analytical form,
at least for the specific choice of the bulk potential under consideration.

In classical dilatonic gravity, a variety of brane-world solutions have
been presented in \Ref {CEGH} where the problem of singularities was also
discussed. A fine-tuned example of bulk potential where one gets a
nonsingular bulk solution, has been presented.

Let us find out if our solutions contain curvature singularities.
Multiplying $g_{(5)}^{\mu\nu}$ with the Einstein equation in the bulk,
\bearr
    R_{(5)\mu\nu}-{1 \over 2}\nabla_\mu\phi\nabla_\nu\phi
            -{1 \over 2}g_{(5)\mu\nu}\Bigl(R_{(5)}
\nnn \cm\cm               \label{Cur1}
    - {1 \over 2}\nabla_\rho\phi\nabla^\rho\phi
            + {12 \over l^2} + \Phi(\phi)\Bigr) ,
\ear
which is obtained from $\SEH$, one gets
\beq                                                       \label{Cur2}
    R_{(5)} = {1 \over 2}\nabla_\rho\phi\nabla^\rho\phi
    - {5 \over 3}\biggl( {12 \over l^2} + \Phi(\phi)\biggr) .
\eeq
Substituting the expressions (\ref{assmp1}) and (\ref{assmp2})
into (\ref{Cur2}), we find
\beq                                        \label{Cur3}
    R_{(5)} = {p_1^2 \over 2y^2 f} - {5 \over 3}
        \left[ c_1(p_2y)^{ap_1} + c_2(p_2y)^{2ap_1}\right] .
\eeq
Then, for cases 1 $\div$ 3, the scalar curvature $R_{(5)}$ is
\bea                                                       \label{Curc1}
    &\mbox{case 1:}& \quad R_{(5)}
            = - {3 \over 2}{p_1^2 q^2 \over y}, \\
\label{Curc2}
    &\mbox{case 2:}& \quad R_{(5)}
           = {8k \over y} - {2\tilde c_2 \over 3y^2}, \\
\label{Curc3}
    &\mbox{case 3}& \quad R_{(5)}
        = - {33\tilde c_1 \over 21\sqrt{y}} - {4k \over y}.
\eea
In all cases the singularity appears at $y=0$.

In case 1, when $y\sim 0$ and the coordinates besides $y$ are
fixed, the infinitesimally small distance $ds$ is given by
\be                                                       \label{Cur4}
    ds=\sqrt{f}dy \sim {dy \over q\sqrt{y}},
\ee
which tells us that the distance between the brane and the singularity is
finite. Then in the cases $k=0$ and $k<0$ the singularity is naked when we
Wick-re-rotate the space-time to Lorentzian signature. When $k>0$, the
singularity is not precisely naked after the Wick re-rotation since the
horizon is situated $y=0$, i.e., the horizon coincides with the curvature
singularity.

\Acknow{The work has been supported by RFBR.}

\end{document}